\documentclass[12pt,manuscript]{aastex}

\shorttitle{Signatures of impulsive localized heating in the temperature distribution of multi-stranded coronal loops}
\shortauthors{Susino et al.}

\begin{document}
\title{Signatures of impulsive localized heating in the temperature distribution of multi-stranded coronal loops}
\author{R. Susino\altaffilmark{1,2}, A. C. Lanzafame\altaffilmark{1,2}, A. F. Lanza\altaffilmark{2}, and D. Spadaro\altaffilmark{2}}

\altaffiltext{1}{Dipartimento di Fisica e Astronomia - Sezione Astrofisica, Universit\`a di Catania, via S. Sofia 78, I-95123 Catania, Italy}
\altaffiltext{2}{INAF-Osservatorio Astrofisico di Catania, via S. Sofia 78, I-95123 Catania, Italy}

\begin{abstract}
We study the signatures of coronal heating on the differential emission measure (DEM) by means of hydrodynamic simulations capable of resolving the chromospheric-corona transition region sections of multi-stranded coronal loops and following their evolution.
We consider heating either uniformly distributed along the loop or localized close to the chromospheric footpoints, in both steady and impulsive regimes.
Our simulations show that condensation at the top of the loop forms when the impulsive heating, with a pulse cadence
lower than the plasma cooling time, is localized at the loop footpoints and the pulse energy is below a threshold above which the heating balances the radiative losses, thus preventing the catastrophic cooling which triggers the condensation.
A condensation does not produce observable signatures in the DEM because it does not redistribute the plasma over a sufficiently large temperature range.
On the other hand, the DEM coronal peak is found sensitive to the pulse cadence time when this is longer or comparable to the plasma cooling time. In this case, the heating pulses produce large oscillations in temperature in the bulk of the coronal plasma, which effectively smears out the coronal DEM structure.
The pronounced DEM peak observed in active regions would indicate a predominance of conditions in which the cadence time is shorter or of the order of the plasma cooling time, whilst the structure of the quiet Sun DEM suggests a cadence time longer than the plasma cooling time.
Our simulations give an explanation of the warm overdense and hot underdense loops observed by TRACE, SOHO and \textit{Yohkoh}.
However, they are unable to reproduce both the transition region and the coronal DEM structure with a unique set of parameters, which outlines the need for a more realistic description of the transition region.
\end{abstract}


\section{Introduction}\label{sec:intro}
Solving the problem of the heating of the solar corona is one of the major issues in solar physics.
In particular, the question whether the plasma heating inside coronal structures is the effect of steady or impulsive processes, uniform or localized within the structure, is still open.

Since the earliest observations, the solar corona appeared as composed by discrete bright structures, called coronal loops, consisting of magnetic flux tubes in which the hot and dense plasma is confined by the magnetic field.
Coronal loops are characterized by different lengths, temperatures, activity levels, and appear to evolve with lifetimes of the order of several hours.
The first models of loop heating \citep[e.g.,][]{ros78,ser81} considered flux tubes of constant cross-section filled with plasma in hydrostatic equilibrium, and in energy balance under the effects of uniform steady heating, conductive heat flux, and radiative losses.
These models predict scaling laws relating the temperature, density, and length of a loop, and allow us to reproduce quite satisfactorily the X-rays coronal emission of both the Sun and solar-type stars.

Nevertheless, recent TRACE and SOHO observations have provided evidence that a large majority of warm coronal loops ($T\sim 1$--2~MK), although appearing in quasi-static conditions, are indeed over dense \citep{asc99,asc01,win03,pat04}, while hot loops (\mbox{$T>2$}~MK) observed by \textit{Yohkoh} exhibit lower densities than predicted by loop models in hydrostatic equilibrium \citep{por95}.
These discrepancies could be explained if coronal loops are assumed to consist of unresolved magnetic strands, each of them heated impulsively and non-uniformly, at different times from its neighbors \citep{car94,kli01,spa03,car04,rea05,pat05,kli06,kli08}.
The idea of impulsive coronal heating was firstly proposed by \citet{par83,par88}, who introduced the concept of nanoflare, a local small-scale ($\lesssim 100$~km) event in which roughly $10^{24}$~erg of energy are released in the corona by magnetic field reconnection, after the magnetic stresses induced on the flux tubes by photospheric motions have reached a certain threshold.

In this Paper we investigate the response of magnetic loop plasma to different kinds of energy deposition by performing numerical simulations with a code capable of resolving the chromospheric-coronal transition region sections of the loop and following them as they respond to heating variations.
We consider heating either uniformly distributed along the loop or localized close to the chromospheric footpoints, in both steady-state and impulsive regimes.
In the impulsive case, we consider different values for the cadence of the injection of the energy pulses into the coronal segment of the loop, in order to perform a first comprehensive study of the consequences of the variation of such a parameter for the loop structure and evolution.
The hydrodynamic behavior of coronal loops undergoing different heating regimes and the relevant consequences on the
differential emission measure are discussed in detail, together with the indications that the variety of conditions found in this exploration give on the physical origins of coronal heating and related phenomena.

\section{Numerical model and simulations}\label{sect:model}
We performed numerical simulations of coronal loop hydrodynamics under different heating conditions using ARGOS \citep{ant99,mac00}, a one-dimensional code that solves the standard set of equations for the conservation of mass, momentum and energy by means of a high-order Godunov scheme and an adaptive mesh refinement.
ARGOS also properly accounts for the heat flux saturation that may occur in some low-density and high-temperature regimes \citep[see, e.g.,][]{kli06}.

We considered a flux tube with a 80~Mm long coronal segment, to model typical active region loops, as observed, for instance, by TRACE.
A detailed description of the characteristic features of the loop model can be found in \citet{spa03}.

The loop was initially settled to a nearly stable equilibrium state under the action of a spatially uniform and temporally constant background heating \citep[as described, e.g, in][]{ant99,ant00}.
A heating rate \mbox{$E_{\rm{base}}=2\times10^{-5}$}~erg~cm$^{-3}$~s$^{-1}$ was chosen, according to the canonical hydrostatic scaling laws \citep[see][]{ros78}, to get an apex temperature of about 0.75~MK at the end of the relaxation phase.
At this stage, the loop structure is that of a steady loop as in \citet{ros78}.
The uniform background heating applied to achieve the initial equilibrium was turned off, within the coronal segment of the loop, when we started the additional heating ($t=0$) and remained off for the remaining of the simulation.
In the chromospheric part of the loop, the $E_{\rm{base}}$ heating was maintained throughout the simulation, although it has a little influence on the loop hydrodynamics, due to the high density in the chromosphere.

The additional heating rate, $E(s,t)$, is assumed to be a separable function of the curvilinear coordinate along the field lines, $s$, and of the time, $t$:
\begin{equation}
  E(s,t)=F(s)\cdot G(t)\ \left[\mbox{erg}\ \mbox{cm}^{-3}\ \mbox{s}^{-1}\right].
\end{equation}
We examined all the possible combinations among two kinds of heating spatial distribution, i.e. quasi-uniform\footnote{Note that in the \textit{quasi-uniform case} the chromospheric and coronal heating rates are different.} or localized at the loop footpoints, and two kinds of temporal distribution, i.e. impulsive or steady.

In the localized heating cases, the heating rate has a maximum at the loop footpoint, and exponentially falls off in the corona with a fixed scale-length of 10~Mm; the location and scale-length of the energy deposition are consistent with those deduced from TRACE observations \citep{asc00,asc01}.
We also consider the possibility of an asymmetric energy deposition between the two footpoints of the loop, since real coronal loops do not appear to be symmetric.
For the impulsive heating cases, we inject into the coronal segment of the loop a sequence of energy pulses, or nanoflares, with constant cadence, duration, and energy amplitude; we model each nanoflare with a Maxwellian function, such that the energy release $E(s,t)$ during a single event has a steep rise, followed by a more gradual decrease.
Specifically:
\begin{equation}
  \begin{array}{ccll}
    F(s) & = & {\left\{
      \begin{array}{ll}
	f \exp \left(-\frac{s-s_1}{\lambda} \right), s\geq s_1 \\
	1
      \end{array}
      \right.} &
    \begin{array}{ll}
      \mbox{localized heating} \\ \mbox{quasi-uniform heating}
    \end{array} \\
    G(t) & = & {\left\{
      \begin{array}{ll}
	E_{I} \frac{1}{2 \tau^3} t^2 \exp \left(-\frac{t}{\tau} \right) \\
	E_{S}
      \end{array}
      \right.} &
    \begin{array}{ll}
      \mbox{impulsive heating} \\ \mbox{steady heating}
    \end{array}
  \end{array}
\end{equation}
where $f=0.75$ or 1.0 measures the constant ratio of the localized heating at the right footpoint to that at the left one, $s_1$ is the initial position of the top of the chromosphere at each footpoint, $\lambda=10$~Mm is the scale-length of the heating distribution, $E_I$ gives the volumetric heating per nanoflare event in the impulsive case, $\tau$ is a characteristic parameter related to the duration of the energy pulses (here always equal to 12.5 s), and $E_S$ is the constant heating rate in the steady case.
Table~\ref{tab:1} lists the values of the parameters adopted in the simulations performed for this study.

\begin{table}
  \caption{Parameters of the simulations}
  \label{tab:1}
  \medskip 
  \begin{tabular}{ccccccc}
    \tableline
    Run~& $\lambda$ & $t_c$  & $E_P$\tablenotemark{a} & $E_{max}$\tablenotemark{b} & $<\!E\!>$ \tablenotemark{c} & Class \\
    \#  & (Mm)	    & (s)    & ($10^{24}$~erg)        & (erg~cm$^{-3}$~s$^{-1}$)   & ($10^{21}$~erg s$^{-1}$) \\   
    \tableline
    1  & Quasi-uniform  & Steady & 1                  & $5\times10^{-4}$           & 4   & Steady             \\
    2  & Quasi-uniform  & 250    & 1                  & $2\times10^{-3}$           & 4   & Steady             \\
    3  & Quasi-uniform  & 500    & 1                  & $2\times10^{-3}$           & 2   & Dynamic            \\ 
    4  & Quasi-uniform  & 1000   & 1                  & $2\times10^{-3}$           & 1   & Dynamic            \\
    5  & Quasi-uniform  & 2000   & 1                  & $2\times10^{-3}$           & 0.5 & Dynamic            \\
    6  & 10             & Steady & 1                  & $4\times10^{-3}$           & 4   & Condensation       \\
    7  & 10             & 250    & 0.125              & $2.75\times10^{-3}$        & 0.5 & Condensation       \\
    8  & 10             & 250    & 0.5                & $1.1\times10^{-2}$         & 2   & Condensation       \\
    9  & 10             & 250    & 1                  & $2.2\times10^{-2}$         & 4   & Condensation       \\
    10 & 10             & 250    & 2                  & $4.4\times10^{-2}$         & 8   & Steady             \\
    11 & 10             & 250    & 4                  & $8.8\times10^{-2}$         & 16  & Steady             \\
    12 & 10             & 500    & 0.25               & $5.5\times10^{-3}$         & 0.5 & Dynamic            \\
    13 & 10             & 500    & 1                  & $2.2\times10^{-2}$         & 2   & Dynamic            \\
    14 & 10             & 500    & 2                  & $4.4\times10^{-2}$         & 4   & Dynamic            \\
    15 & 10             & 1000   & 1                  & $2.2\times10^{-2}$	   & 1   & Dynamic            \\
    16 & 10             & 1000   & 4                  & $8.8\times10^{-2}$	   & 4   & Dynamic            \\
    17 & 10             & 2000   & 1                  & $2.2\times10^{-2}$	   & 0.5 & Dynamic            \\
    18 & 10             & 2000   & 8                  & $1.76\times10^{-1}$        & 4   & Dynamic            \\
    \tableline
  \end{tabular}
  \tablenotetext{a}{Total energy per pulse, or, in the steady cases, total energy deposited into the loop in a time interval equal to $t_c=250$~s.}
  \tablenotetext{b}{Maximum heating rate per unit volume.}
  \tablenotetext{c}{Average amount of energy supplied to the loop per unit time.}
\end{table}

We chose the value of $E_I$ so that
\begin{equation}
  \int_{V}{dV}\int_{0}^{t_c}{E(s,t)\,dt}=E_P\ \left[\mbox{erg}\right],
\end{equation}
where $V$ is the volume of the coronal section of the loop, $t_c$ is the cadence time between two consecutive nanoflares, and $E_P$ is the amount of energy supplied by each pulse.
For the steady heating cases, the value of $E_S$ was adjusted so that the total energy deposited into the loop in a time interval $t_c$ was equal to $E_P$, as listed in Table~\ref{tab:1}.
In this case $E_S$ is equal to $E_{max}$, as reported in Table~\ref{tab:1}.

The value of the time interval between the pulses, i.e. the cadence $t_c$, was fixed taking into account the characteristic radiative cooling time of the loop, $\tau_{\rm cool}$.
For a loop with a semilength $L=40$~Mm and an apex temperature $T_{\rm apex}=2$~MK, it
results $\tau_{\rm cool}\approx 1000$~s, according to, e.g., \citet{ser91}.
Here we report results obtained with $t_c$ ranging from $250\ \mbox{s}\approx\tau_{\rm cool}/4$ to $2000\ \mbox{s}\approx 2\,\tau_{\rm cool}$.

We also considered different values for $E_P$, in order to change the amount of energy supplied to the loop by the sequence of heating pulses and investigate the related effects on the plasma hydrodynamics.

\section{Results and discussion}
\subsection{Plasma dynamics}
\begin{figure}
  \centering
  \includegraphics[width=0.80\textwidth]{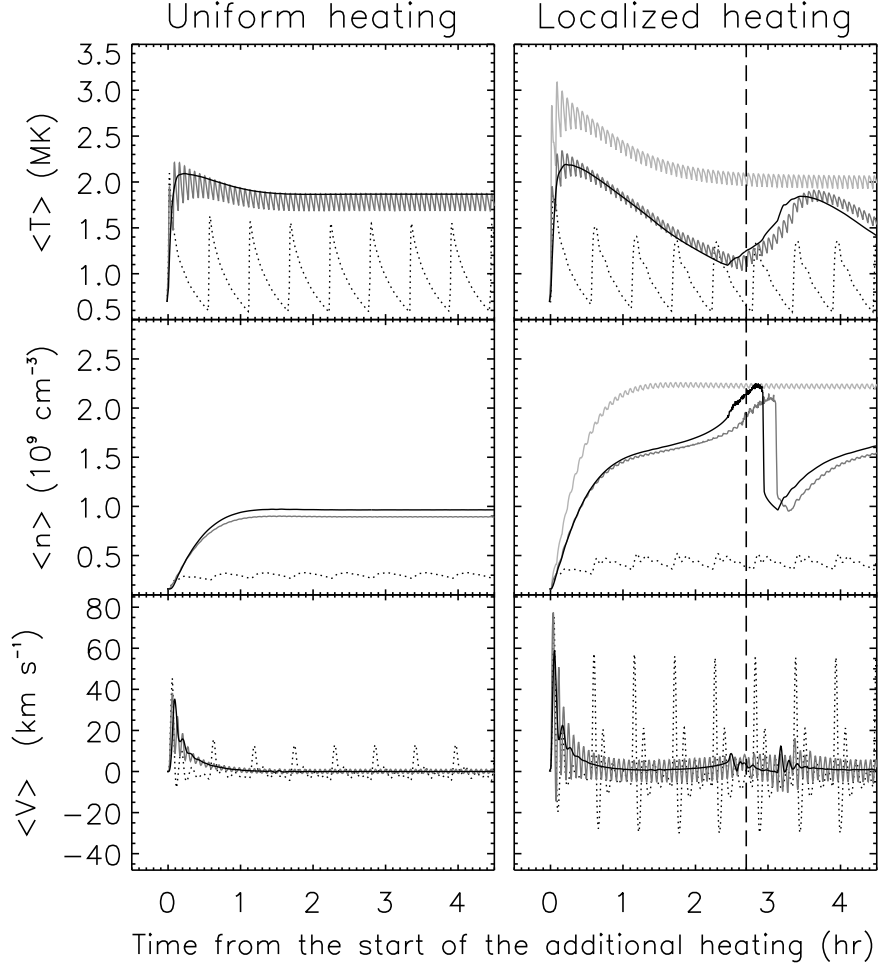}
  \caption{Initial part of the temporal evolution of the model temperature (top panels), density (middle panels), and velocity (bottom panels), averaged over the upper 3/4 of the loop coronal segment, for run~1,~2, and~5 (left panels; black, heavy gray and dotted lines, respectively), and run~6,~9,~10, and~17 (right panels; black, heavy gray, light gray, and dotted lines, respectively).
  The velocity is almost identical in runs~9 and~10 and therefore only the run~9 velocity is reported in the bottom-right panel. The vertical dashed lines in the right panels approximatively correspond to the maximum of the condensation phase for run~9, whose snapshot is given in Fig.~\ref{fig:snapshot}.\label{fig:evol}}
\end{figure}

Figure~\ref{fig:evol} shows the initial part of the temporal evolution of the plasma temperature, density, and velocity averaged over a relevant portion (3/4) of the loop coronal segment \citep[see, e.g.,][]{pat05} in some representative cases.

In the impulsive quasi-uniform heating case with $E_P=10^{24}$~erg and $t_c=250$~s (run~2), the loop model simply settles into a new quasi-static equilibrium state, characterized by nearly constant temperature, density and velocity, apart from small oscillations, lower then 10\%, due to the sequence of energy pulses.
Note that the average values of these parameters behave as the corresponding ones obtained for the steady quasi-uniform heating case (run~1).
After the start of the additional heating, the temperature of the loop immediately reaches a value of about 2.3~MK; then, in the next hour, it decreases and stabilizes around a value slightly lower than 2~MK.
The density monotonically increases from $1.4\times10^{8}$~cm$^{-3}$ to $\sim9.0\times10^{8}$~cm$^{-3}$, while the plasma velocity, after a first highly dynamic phase caused by the abrupt increase of the heating rate and characterized by upflow velocities up to 40~km~s$^{-1}$, approaches very low values of some km~s$^{-1}$.
The time-averaged plasma temperature and density for this loop model are consistent with the hydrostatic scaling laws defined by \citet{ros78}.

\begin{figure}
  \centering
  \includegraphics[width=0.80\textwidth]{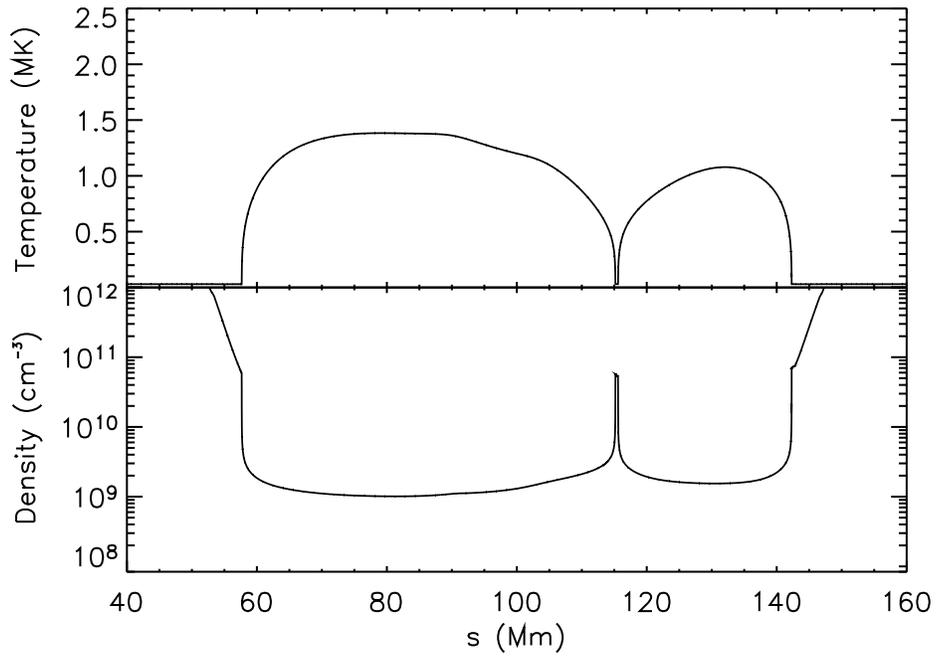}
  \caption{Instantaneous profiles of the loop model temperature (upper panel) and density (lower panel) vs. the curvilinear coordinates for run~9, at 2.7~hours after the start of the additional heating, marked with a vertical line in Fig.~\ref{fig:evol}.
  The chromospheric segments have been truncated to highlight the coronal portion of the loop.\label{fig:snapshot}}
\end{figure}

Conversely, asymmetric heating localized near the loop footpoints, both steady (run~6) and impulsive with $t_c=250$~s (run~9), causes a more dynamic evolution for similar values of the heating parameters, with long-term fluctuations of temperature and density (after an initial steep increase) due to cycles of plasma condensation formation, motion along the loop, and falling onto the nearest and less heated footpoint (note that the first cycle only is reported in Fig.~\ref{fig:evol}).

The phenomenon of plasma condensation formation is well known and extensively described in the literature \citep[e.g.,][]{ant99,ant00,kar01,kar06,kar08,kli09,mul03,mul04,tes05}.
Recently, it has also been observed in coronal non-flaring loops \citep[e.g.,][]{osh07}.
It is the effect of a thermal instability occurring near the top of the loop, where the energy supply is not sufficient to balance the radiative losses induced by the increase of the plasma density owing to the strong chromospheric evaporation.
This eventually leads to a catastrophic cooling, down to chromospheric temperatures, of the material located close to the loop apex and, consequently, to the formation of a region of low-temperature and high-density plasma.
This is clearly visible in Fig.~\ref{fig:snapshot}, which shows the instantaneous plasma temperature and density profiles along the loop (run~9) at a stage of its evolution characterized by the presence of a condensation.
The blob of cool and dense plasma subsequently starts moving slowly toward the less heated footpoint, because of the differences in pressure induced by the asymmetric energy release in the two legs of the loop, and finally drains onto the chromosphere.
The depleted loop then reheats quickly and a new cycle of chromospheric evaporation, plasma condensation and draining takes place.
In the present model the first catastrophic cooling phase approximatively begins 2.5~hours after the start of the additional heating and lasts for about one hour; its total duration slightly depends on whether the energy deposition is impulsive or steady, in the latter case being nearly 20~minutes shorter.
The periodicity of the described cycle is about three hours.

The temporal evolution of run~7 and~8 (with $E_P<10^{24}$~erg, not shown in Fig.~\ref{fig:evol}) is very similar to that of run~9, with, however, lower temperatures, densities, and velocities, but again with the onset of a dynamic cycle of plasma condensation formation.

Therefore, the impulsive heating cases, both quasi-uniform (run~2) and localized (run~7,~8, and~9), do not exhibit appreciable differences with respect to the corresponding steady cases (run~1 and~6, respectively).
This is expected when the cadence of nanoflares is very rapid compared to the loop plasma cooling time \citep[see, e.g.,][]{wal97,tes05,kli06}.
In our cases a cadence time equal to $\tau_{\rm cool}/4$ already gives a nearly steady heating situation.
On the other hand, it appears evident that the localization of the heating near the loop footpoints plays a fundamental role in the plasma condensation formation when the energy deposition is steady or impulsive with a cadence time well below the loop cooling time.  These results, obtained for a ratio of the heating damping-length ($\lambda$) to the loop semilength equal to 1/4, are in agreement with those of other works modeling footpoint-heated loops in similar conditions \citep[e.g.,][]{mul03,mul04,tes05}.

Run~5 and~17 illustrate both quasi-uniform and localized heating cases in which the cadence time is longer than the loop plasma cooling time ($t_c>\tau_{\rm cool}$).
In this cases (dotted-line curves in Fig.~\ref{fig:evol}) the temperature in the upper part of the loop coronal segment shows pronounced oscillations since there is enough time between pulses for the plasma to cool down and drain downward to the chromospheric region (as confirmed by the downward velocities of the plasma noticed in between the pulses).
Density has small variations, but velocity shows strongly damped oscillations with a rather high peak at the beginning of each energy pulse.
No condensation forms in this case, because the intermittent heating prevents the accumulation of plasma at the loop top, and thus the thermal instability.
A similar behavior is also obtained when $t_c\simeq\tau_{\rm cool}$ (run~4,~15, and~16) or equal to $\tau_{\rm cool}/2$ (run~3,~12,~13, and~14).

Note, however, that although the heating localization and a high frequency pulse cadence are necessary to yield a catastrophic cooling phase during the loop evolution, they are not always sufficient.
The crucial point is the balance between the energy supplied to the loop top by the sequence of heating pulses and the radiative losses of the plasma accumulated therein.
In fact, by sufficiently increasing the amount of energy supplied by each pulse, even without changing the other parameters of the heating regime, such a balance can be achieved, thus preventing the thermal instability and the consequent plasma condensation formation.
This is the case of run~10 (see left panels in Fig.~\ref{fig:evol}) and run~11.

As far as the quasi-uniform loop simulations (runs~1--5) are concerned, it is also worth noting that increasing $t_c$ between runs~2--5 corresponds to a decrease by a factor of~8 in the total average energy $<\!\!E\!\!>$ put into the corona per second, because $E_P$ is held fixed. Since no condensation forms, the average temperature and density are essentially controlled by the mean energy dissipated per unit time, as in the steady case (see also Sect.\,\ref{sec:dem_quasi_uniform_impulsive}).

\begin{figure}
  \centering
  \includegraphics[width=0.60\textwidth]{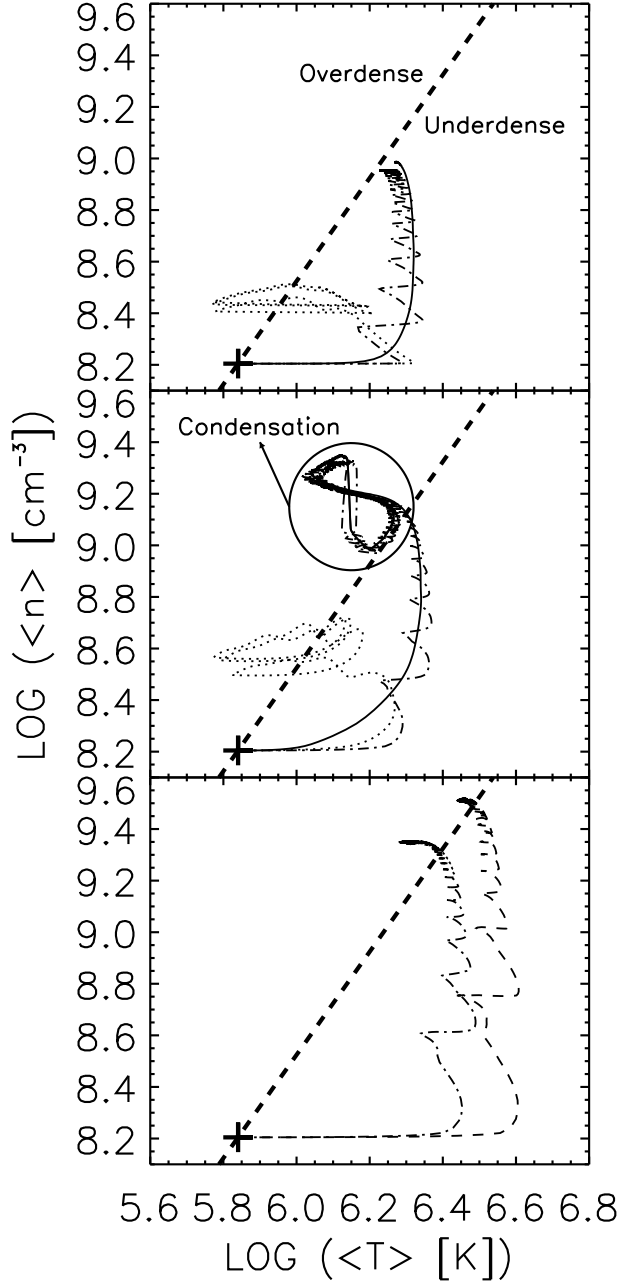}
  \caption{Evolution of the average loop density and temperature for runs~1,~2, and~5 (top panel; solid, dash-dotted, and dotted line, respectively), run~6,~9, and~17 (middle panel; solid, dash-dotted, and dotted line, respectively), and run~10 and~11 (bottom panel; dash-dotted and dashed line, respectively).
  The hydrostatic scaling law \citep[heavy dashed line, see][]{ros78} is also reported, gauged according to the initial static equilibrium (indicated by the cross).
  In the middle panel, the evolution characterized by condensation formation cycles occurs within the encircled region labeled ``condensation''.\label{fig:avgNT}}
\end{figure}

Figure~\ref{fig:avgNT} reports the average loop density and temperature during the evolution of some representative loops together with the hydrostatic scaling law relationship.
We note that all the examined loops exhibit densities significantly smaller than those predicted by the scaling laws in the initial phase of their evolution (for about one hour), in both steady-state and impulsive cases.
Even if the onset of the heating produces an increase in the loop temperature and a large conductive heat flux at the coronal base of the loops, driving the upward expansion known as chromospheric evaporation \citep{a&s78}, there is a time lag in the corresponding increase of the coronal density, so that the loops appear to be hot and underdense.

The subsequent evolution of the loop depends on the characteristics of the plasma heating.

In the quasi-uniform cases (top panel in Fig.~\ref{fig:avgNT}), when the heating is steady (run~1) or impulsive with a cadence time well below the loop cooling time (run~2), the loop evolves toward a new state almost consistent with the hydrostatic scaling laws (only slightly underdense).
When $t_c$ is comparable to or longer than the loop cooling time (e.g. run~5) the loops begin to move along nearly cyclic sequences, spanning from hot, underdense conditions to cool, overdense ones. Note, however, that the range of density values (in dex) is significantly smaller than that covered in temperature.
Moreover, the loops appear significantly overdense only at temperatures well below 1~MK.

In the localized heating cases (middle panel in Fig.~\ref{fig:avgNT}), cadence times comparable to or longer than the loop cooling time (e.g. run~17) give rise to a loop evolution in general agreement with that found in the quasi-uniform cases with similar cadence, although the plots reproducing the sequences exhibit some significant differences, particularly at the higher temperatures.
The hydrodynamic behavior is very similar to that described by \citet{spa03}, who considered localized transient heating on timescales comparable to the loop cooling time.
Note that no condensation forms near the loop apex in these cases.
When the localized heating is steady (run~6) or impulsive with $t_c<\tau_{\rm cool}$ (e.g. run~9), the evolution is considerably different: as the temperature decreases, the plasma continues to evaporate and the density continues to rise, overshooting by a factor of $\sim 5$ the equilibrium values predicted for static loops with the same coronal temperatures, until a plasma condensation forms near the loop apex.
The fall of material on to the less heated loop footpoint causes an abrupt decrease of the average density, although the loop remains slightly overdense and subsequently a new cycle of condensation formation and evolution starts.
Hence these heating conditions can give rise to warm ($T\sim 1$--2~MK), overdense loops with lifetimes of some hours, such as those observed by TRACE and SOHO.

The bottom panel in Fig.~\ref{fig:avgNT} shows that increasing the amount of energy supplied by each pulse, even without changing the other characteristics of the heating regime (e.g. runs~10 and~11), causes the loops to evolve toward a hot ($T\geq 2$--3~MK), quasi-static, slightly overdense state, where they settle after a balance between the energy supplied to the loop top and the radiative losses therein is achieved.
Note that run~11 might apply to the case of the hot ($\simeq 3$~MK) loops seen by the \textit{Yohkoh} Soft X-ray Telescope, without a corresponding warm counterpart ($\simeq 1$~MK) observed by TRACE \citep[see, e.g.,][] {nit00,kli06}.
If these loops are impulsively heated, then the nanoflares must occur frequently enough that the plasma does not have time to cool to TRACE temperatures.

\subsection{Differential Emission Measure}
Comparison of our modeling with observations is done using the differential emission measure, $\mbox{DEM}(T)$ \citep[e.g.,][]{cra76}, which effectively describes the plasma distribution in temperature.

We simulated the multi-strand loop DEM by averaging instantaneous DEMs calculated at $n=300$ different times, randomly selected throughout the simulation.
Although each simulation represents the evolution of a single magnetic strand, we assume that the states of the model at $n$ randomly selected times can be used to describe the behavior of $n$ independent strands observed at the same time, thus giving a single simulated snapshot of a multi-stranded loop \citep[see, e.g.,][]{pat05}.
The theoretical single strand DEM is computed following \citet{pet06}.

In what follows, the observed quiet Sun (QS) and active region (AR) DEMs are adopted from \citet{lan05} (SOHO-CDS observations) and \citet{lan02} (SERTS-89 observations), respectively. 

\subsubsection{Initial conditions and hydrostatic models}
\begin{figure}
  \centering
  \includegraphics[width=0.80\textwidth]{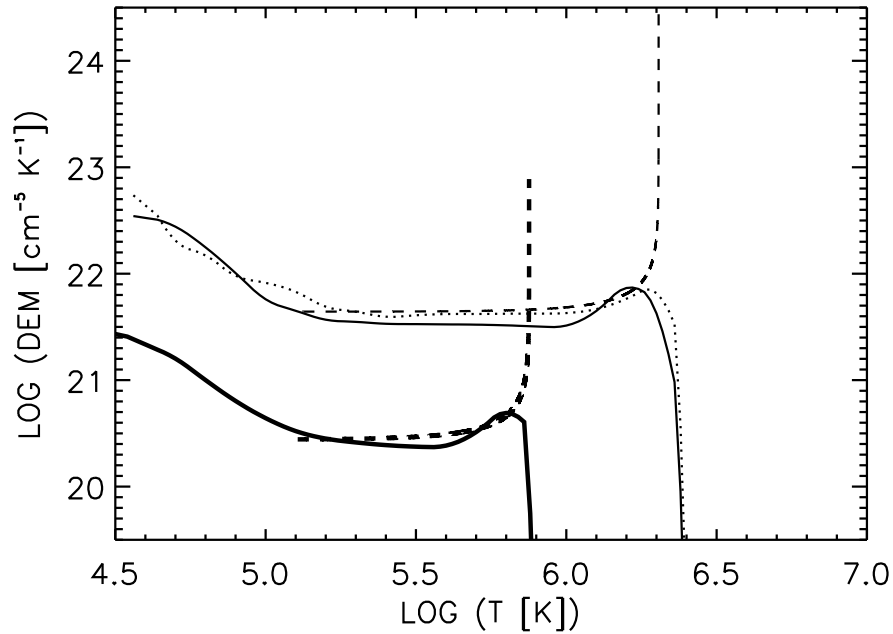}
  \caption{Differential emission measure (DEM) for: the initial steady equilibrium conditions adopted in our simulations with $T_{\rm apex}\simeq 0.75$~MK (thick solid line); the case of a spatially uniform and temporally constant background heating producing a steady loop with an apex temperature of $\simeq 2$~MK at the end of the relaxation phase (dotted line); the quasi-uniform, steady case with $<\!E\!>=4\times 10^{21}$~erg~s$^{-1}$ (run~1, thin solid line).
  For comparison, the analytical DEMs according to the hydrostatic scaling laws \citep[see Eq.~C4 in][]{ros78} are also shown (heavy- and light-dashed lines, respectively).
  \label{fig:initial_RTV}}
\end{figure}

In Fig.~\ref{fig:initial_RTV} we plot the DEM corresponding to: i) the initial steady equilibrium conditions adopted in our simulations ($T_{\rm apex}\simeq 0.75$~MK, see Sect.~\ref{sect:model}); ii) the case of a spatially uniform and temporally constant background heating with $E_{\rm base}=5\times 10^{-4}$~erg~cm$^{-3}$~s$^{-1}$ producing a steady loop with an apex temperature of about 2~MK at the end of the relaxation phase described in Sect.~\ref{sect:model}; iii) the quasi-uniform, steady case with $<\!E\!>=4\times 10^{21}$~erg~s$^{-1}$ (run~1).
For comparison, the analytical DEMs computed according to the hydrostatic equilibrium laws \citep{ros78} are also shown. The divergence of the analytical results close to the top of the loop comes from the DEM formula, which is written in terms of the inverse of a temperature gradient that vanishes at the loop maximum temperature.
The figure shows that our quasi-uniform, steady simulation for a multi-strand reproduces quite closely the steady equilibrium single loop structure obtained at the end of the relaxation phase with a spatially uniform and temporally constant background heating of the same level.
Apex temperatures and DEM minima at the end of the relaxation phase and in the quasi-uniform, steady simulations are essentially the same as those calculated using the \citet{ros78} relationships.

\subsubsection{Quasi-uniform impulsive cases}
\label{sec:dem_quasi_uniform_impulsive}
\begin{figure}
  \centering
  \includegraphics[width=0.80\textwidth]{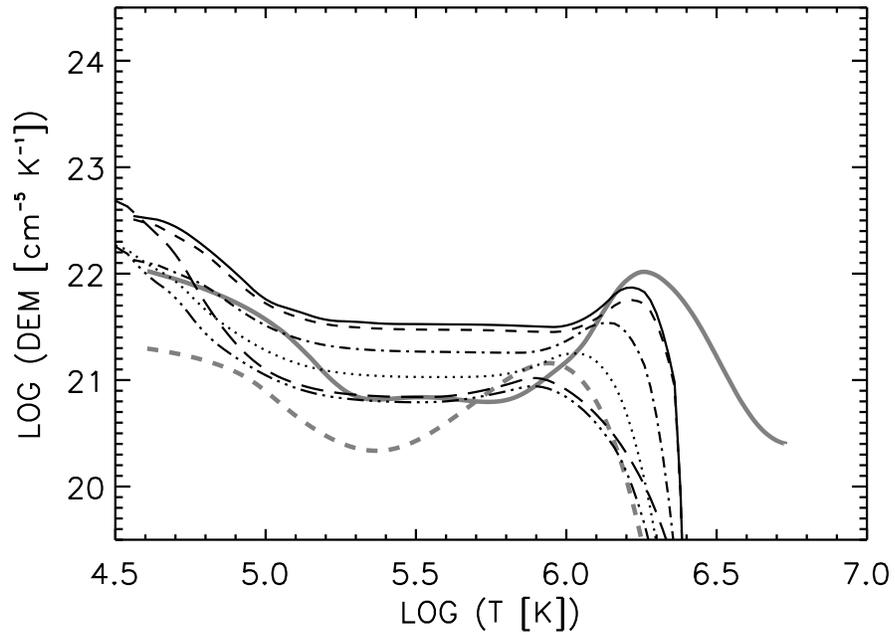}
  \caption{DEM from quasi-uniform impulsive simulations: run~2 (dashed line); run~3 (dash-dotted line); run~4 (dotted line); run~5 (dash-triple-dotted line).
  Comparison is done with QS (gray thick dashed line), AR (gray thick solid line), the quasi-uniform steady case (run~1, solid line), and a single pulse simulation (long-dashed line).\label{fig:dem_uniform_runs}}
\end{figure}

DEMs obtained from our quasi-uniform impulsive simulations are shown in Fig.~\ref{fig:dem_uniform_runs}.
Since no condensation forms, the DEM is controlled essentially by the mean energy dissipated per unit time.
The runs reported in Fig.~\ref{fig:dem_uniform_runs} have fixed $E_P=10^{24}$~erg, and therefore the shorter $t_c$, the higher $<\!E\!>$.
The $<\!E\!>$ values chosen in Fig.~\ref{fig:dem_uniform_runs} correspond to cases in which the models reproduce either the observed AR DEM around $\log T=5.5$ (and the QS DEM peak at $\log T\approx 6$), or the AR DEM temperature peak at $\log T\approx 6.3$, plus a couple of cases with energies in between.
Also shown in Fig.~\ref{fig:dem_uniform_runs} a simulation with a single pulse with $E_P=10^{24}$~erg, which reproduces very closely the conditions explored by \citet{kli08}.
From the comparison with the observed DEMs, we see that these models are unable to reproduce both the TR and coronal DEM, neither for the QS nor for the AR.

The quasi-uniform impulsive model with the highest mean energy dissipated per unit time (run~2, $t_c=250$~s, $<\!E\!>=4\times 10^{21}$~erg~s$^{-1}$) is compared with the steady-uniform model with the same mean dissipated energy rate (run~1), confirming that there is no appreciable difference between the two.

Note also that by increasing $t_c$ well above $\tau_{\rm cool}$ (e.g. run~5 with $t_c=2000$~s), we obtain a DEM structure very similar to that presented by \citet{kli08}.
A close comparison with their simulations has been made by calculating the DEM for a single pulse and spatially uniform heating.
The resulting DEM, also shown in Fig.~\ref{fig:dem_uniform_runs}, is almost identical, above $\log T=5.5$, to those in Fig.~2 of \citet{kli08}.

\subsubsection{Localized impulsive models}
\begin{figure}
  \centering
  \includegraphics[width=0.80\textwidth]{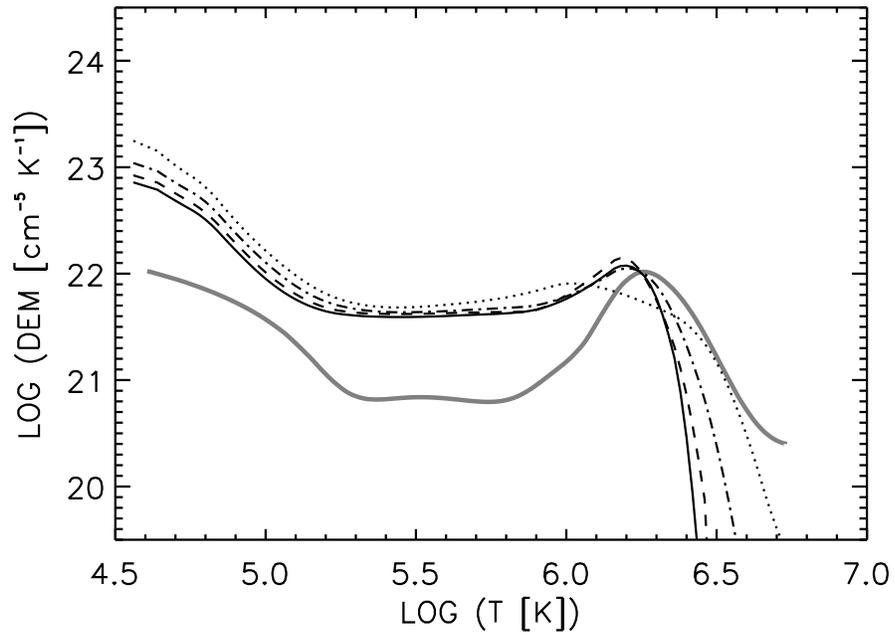}
  \caption{DEM for impulsive localized models with different combinations of $t_c$ and $E_P$ giving the same mean energy dissipated per unit time $<\!E\!>=4\times 10^{21}$~erg~s$^{-1}$: run~9 (solid line); run~14 (dashed line); run~16 (dash-dotted line); run~18 (dotted line).
  Comparison is made with the AR DEM (gray thick solid line).\label{fig:same_energy_ar}}
\end{figure}

In Fig.~\ref{fig:same_energy_ar} we consider DEMs obtained for impulsive localized models with different combinations of $t_c$ and $E_P$ giving the same mean energy rate.
Choosing $<\!E\!>=4\times 10^{21}$~erg~s$^{-1}$, we obtain a DEM peak similar to that observed in the AR.

The mean dissipated energy determines the temperature at which the DEM is at its maximum ($T_{\rm peak}$).
Increasing $t_c$ up to $t_c\approx\tau_{\rm cool}$ causes an increase in the DEM above $T_{\rm peak}$ and a little change in the DEM slope below $T_{\rm peak}$.
Increasing further $t_c$ ($t_c>\tau_{\rm cool}$), causes a further increase in the DEM in the high temperature range (above $\log T\approx 6.5$ in the cases shown in Fig.~\ref{fig:same_energy_ar}) and the disappearance of the coronal DEM peak.

The differences between the cases in which condensation occurs (e.g. run~9) or does not occur (e.g. run~14) are too small to be appreciated in practice.
Also, the localized steady model DEM (run~6) is almost identical to that of the localized impulsive model with the same $<\!E\!>$ and $t_c=250$~s~$<\tau_{\rm cool}$.
Condensation does not produce appreciable differences in the DEM because the bulk of the plasma maintains a temperature above $\sim 1$~MK throughout the evolution (compare with Figs.~\ref{fig:avgNT} and~\ref{fig:evol}).
The sensitivity to $t_c$, on the other hand, derives from the fact that, when $t_c$ increases well above $\tau_{\rm cool}$, condensation does not occur anymore and the bulk of the plasma has a temperature oscillating over a rather large range.
In our simulations, such oscillation produces a smearing of the DEM around $\sim 1$~MK and eventually leads to the disappearing of the DEM peak.

Also in this case, the simulations are unable to reproduce the whole observed DEM structure.
By appropriately selecting $t_c$ and $E_P$, it is possible to reproduce the DEM in the high temperature range or around the minimum DEM, but it is not possible to reproduce the DEM in both ranges with a single ($t_c$, $E_P$) pair.
We conclude, therefore, that the assumption of multi-strand structures subject to localized impulsive heating is not sufficient to explain the well known discrepancies with observations, which must be due to physical processes not included in our simulation.
The recent suggestion put forward by \citet{jud08} of cross-field diffusion of neutral atoms from cool threads extending into the corona may help in solving problems like this. Spicular absorption of some of the plasma emission below 1\,MK, as suggested by \cite{kli08} and \cite{dep09}, might also contribute to explain the discrepancies with the observed DEM structure.

Nevertheless, we suggest that, despite such discrepancies, the systematic behavior of the simulated DEM with $t_c$ would maintain its validity even if the models are not sufficiently detailed, and would help in discriminating at least among very different conditions.
In fact, despite the small sensitivity of the DEM to variations in $t_c$, when this increases well above $\tau_{\rm cool}$, the coronal DEM peak tends to disappear.
Such changes in the DEM shape can be reliably verified by comparison with observations, since they are well above
the expected uncertainties in the DEM reconstruction.
The very existence of the coronal DEM peak, therefore, is an indication of the existence of heating pulse cadence time shorter than $\tau_{\rm cool}$.

\begin{figure}
  \centering
  \includegraphics[width=0.80\textwidth]{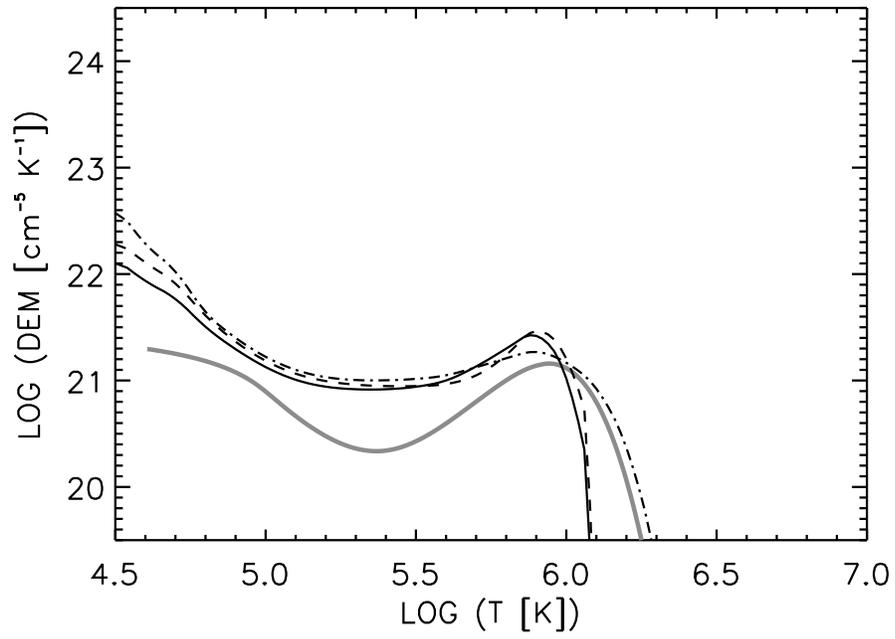}
  \caption{DEM for simulations with $<\!E\!> = 5 \times 10^{20}$~erg~s$^{-1}$ (run~7, solid line; run~12, dashed line; run~17, dash-dotted line) compared with the QS DEM (gray thick solid line).\label{fig:same_energy_qs}}
\end{figure}

Finally, in Fig.~\ref{fig:same_energy_qs} we show the simulations with $<\!E\!>=5\times 10^{20}$~erg~s$^{-1}$ (runs~7,~12, and~17) compared with the observed QS DEM.
The dependence on $t_c$ described above for the AR DEM is maintained at such lower energy, which corresponds to a DEM peak close to the observed QS one.
The DEM peak decreases at increasing $t_c$, tending to disappear for $t_c>\tau_{\rm cool}$.
The DEM at temperatures above $T_{\rm peak}$ increases with increasing $t_c$.

Comparing the simulated DEM's peaks with AR and QS observations (Figs.~\ref{fig:same_energy_ar} and~\ref{fig:same_energy_qs}), the pronounced peak in the observed AR DEM may indicate a predominance of $t_c<\tau_{\rm cool}$ conditions.
On the contrary, the less pronounced peak in the observed QS DEM may indicate that $t_c>\tau_{\rm cool}$ conditions dominate in that regime.
Although the shape of the observed DEM may be affected by the smoothing imposed by the regularization technique used to reconstruct the DEM from the observed spectral line intensities, the differences implied by the models are higher than the uncertainties in the DEM reconstruction due to regularization smoothing and intensity ratios of spectral lines formed around and above the DEM peak could be useful to discriminate among the high- and low-cadence regimes.

It is also worth noting that, despite a close fit to the observed DEM is outside the scope of this work and only a general comparison is made, models that reproduce more closely the AR DEM have a coronal electron pressure $\log P_{\rm e} \sim 15.3$ ($P_{\rm e}$ in units cm$^{-3}$\,K) remarkably close to the value deducted by \cite{lan02} using line-ratio diagnostics ($\log P_{\rm e} \sim 15.8$). In the QS case, models that reproduce more closely the observed DEM have $\log P_{\rm e} \sim 14.0$ in the coronal part, in close agreement with the \cite{lan05} estimate ($\log P_{\rm e} \sim$ 13.7 -- 14.0).

\section{Conclusions}
We have carried out hydrodynamic simulations of multi-stranded coronal loops with a code capable of resolving the transition region sections and following their evolution in order to outline signatures of the heating regimes in the observed DEM.

Quasi-uniform or localized impulsive heating with a cadence time $\simeq 1/4$ of the plasma cooling time can produce a plasma evolution which is essentially indistinguishable from the corresponding steady cases.

Plasma condensation occurs on a limited range of heating parameters.
Energy localization is necessary to yield a catastrophic cooling phase during the loop evolution, but it is not always sufficient.
The crucial point is the balance between the energy supplied to the loop top and the radiative losses therein.
Thus, the variation of the energy deposition parameters (such as the pulse cadence, the pulse energy, or the heating damping-length) with respect to the global characteristics of the model (the radiative cooling time or the loop length) could, in some cases, prevent the occurrence of a dynamic cycle of plasma condensation formation even in the presence of a localized heating.
This may happen, for instance, by increasing the heating rate at the loop footpoints or the ratio of the heating damping-length to the loop length, or considering nanoflare cadence times comparable to or longer than the characteristic radiative cooling time.

The DEM is found insensitive to the presence of condensation because the sequence of catastrophic cooling, draining toward the loop footpoint, reheating and evaporation does not effectively redistribute the plasma over temperature and the global distribution remains very close to the corresponding steady configuration.
On the contrary, pulses with cadence longer than the plasma cooling time produce temperature oscillations in the bulk of the plasma which effectively smear the coronal DEM structure.
The effects are observable since the coronal DEM peak tends to disappear when the pulse cadence is about 2 times the plasma cooling time.
The pronounced DEM peak observed in active regions would indicate a predominance of conditions in which the cadence time is shorter or of the order of the plasma cooling time, whilst the structure of the quiet Sun DEM suggests a cadence time longer than the plasma cooling time.

The warm overdense and hot underdense loops observed by TRACE, SOHO \citep{asc99,asc01,win03,pat04} and \textit{Yohkoh} \citep{por95} could be explained by the dynamic evolution of the plasma.
In particular, a localized heating producing plasma condensation cycles can give rise to warm ($T\simeq 2$~MK), overdense loops with lifetimes of some hours, such as those observed by TRACE and SOHO.
Moreover, increasing the amount of energy supplied by each pulse at the loop footpoints prevents the plasma condensation formation and causes the loops to settle in a hot ($T\simeq 2$--3~MK), quasi-static, slightly overdense state, which can reproduce the case of hot loops seen by \textit{Yohkoh} SXT, without a corresponding warm counterpart observed by TRACE \citep[e.g.,][]{nit00}.

Nevertheless, our simulations are unable to reproduce both the transition region and the coronal DEM structure with a unique set of parameters, which suggest that some additional physical processes, like that proposed by \citet{jud08} or by \cite{kli08} and \cite{dep09}, must be taking place in the transition region.

\acknowledgments
DS acknowledges useful discussions with the members of the ISSI team ``The role of the Spectroscopy and Imaging Data in Understanding Coronal Heating" (team Parenti).

This work was supported in part by the Agenzia Spaziale Italiana (contract I/015/07/0 and agreement ASI/INAF I/023/09/0).
Financial support by the European Commission through the SOLAIRE Network (MTRN-CT-2006-035484) is also gratefully acknowledged.


\begin{thebibliography}{}
	\bibitem[Antiochos et al.(1999)]        {ant99} Antiochos, S. K., MacNeice, P. J., Spicer, D. S., \& Klimchuk, J. A. 1999, \apj, 512, 985
	\bibitem[Antiochos et al.(2000)]        {ant00} Antiochos, S. K., MacNeice, P. J., \& Spicer, D. S. 2000, \apj, 536, 494
	\bibitem[Antiochos \& Sturrock(1978)]	  {a&s78} Antiochos, S. K., \& Sturrock, P A. 1978,\apj, 220, 1137
	\bibitem[Aschwanden et al.(1999)]       {asc99} Aschwanden, M. J., Newmark, J. S., Delaboudini\`ere, J.-P., Neupart, W. M., Klimchuk, J. A., Gary, G. A. et al. 1999, \apj, 515, 842
	\bibitem[Aschwanden et al.(2000)]       {asc00} Aschwanden, M. J., Nightingale, R. W., \& Alexander, D. 2000, \apj, 541, 1059
	\bibitem[Aschwanden et al.(2001)]       {asc01} Aschwanden, M., Schrijver, C. J., \& Alexander, D. 2001, \apj, 550, 1036
	\bibitem[Cargill(1994)]                 {car94} Cargill, P. J. 1994, \apj, 422, 381
	\bibitem[Cargill \& Klimchuk(2004)]     {car04} Cargill, P. J. \& Klimchuk, J. A. 2004, \apj, 605, 911
	\bibitem[Craig \& Brown(1976)]          {cra76} Craig, I. J. D. \& Brown, J. C. 1976, \aap, 49, 239
	\bibitem[De Pontieu et al.(2009)]	{dep09} De Pontieu, B., Hansteen, V. H., McIntosh, S. W., \& Patsourakos, S. \apj, 702, 1016.
	\bibitem[Judge(2008)]                   {jud08} Judge, P. 2008, \apj, 683, L87
	\bibitem[Karpen et al.(2001)]           {kar01} Karpen, J. T., Antiochos, S. K., Hohensee, M., Klimchuk, J. A., \& MacNeice, P. J. 2001, \apj, 553, L85 
	\bibitem[Karpen et al.(2006)]           {kar06} Karpen, J. T., Antiochos, S. K., \& Klimchuk, J. A. 2006, \apj, 637, 531
	\bibitem[Karpen \& Antiochos(2008)]     {kar08} Karpen, J. T. \& Antiochos, S. K. 2008, \apj, 676, 658
	\bibitem[Klimchuk(2006)]                {kli06} Klimchuk, J. A. 2006, Sol. Phys., 234, 41
	\bibitem[Klimchuk \& Cargill(2001)]     {kli01} Klimchuk, J. A. \& Cargill, P. J. 2001, \apj, 553, 440	
	\bibitem[Klimchuk et al.(2009)]         {kli09} Klimchuk, J. A., Karpen, J. T., \& Antiochos, S. K. 2009, \apj, submitted
	\bibitem[Klimchuk et al.(2008)]         {kli08} Klimchuk, J. A., Patsourakos, S., \& Cargill, P. J. 2008, \apj, 682, 1351
	\bibitem[Lanzafame et al.(2002)]        {lan02} Lanzafame, A. C., Brooks, D. H., Lang, J., Summers, H. P., Thomas, R. J., \& Thompson, A. M. 2002, \aap, 384, 242
	\bibitem[Lanzafame et al.(2005)]        {lan05} Lanzafame, A. C., Brooks, D. H., \& Lang, J. 2005, \aap, 432, 1063
	\bibitem[MacNeice et al.(2000)]         {mac00} MacNeice, P. J., Olson, K. M., Mobarry, C., de Fainchtein, R., \& Packer, C. 2000, Comput. Phys. Commun., 126, 330
	\bibitem[M\"uller et al.(2003)]         {mul03} M\"uller, D. A. N., Hansteen, V. H., \& Peter, H. 2003, \aap, 411, 605
	\bibitem[M\"uller et al.(2004)]         {mul04} M\"uller, D. A. N., Peter, \& H., Hansteen, V. H. 2004, \aap, 424, 289
	\bibitem[Nitta(2000)]			              {nit00} Nitta, N. 2000, \solphys, 195, 123
	\bibitem[O'Shea et al.(2007)]           {osh07} O'Shea, E., Banerjee, D., \& Doyle, J. G. 2007, \aap, 475, L25
	\bibitem[Parker(1983)]                  {par83} Parker, E. N. 1983, \apj, 264, 642
	\bibitem[Parker(1988)]                  {par88} Parker, E. N. 1988, \apj, 330, 474
	\bibitem[Patsourakos et al.(2004)]      {pat04} Patsourakos, S., Klimchuk, J. A., \& MacNeice, P. J. 2004, \apj, 603, 322
	\bibitem[Patsourakos \& Klimchuk(2005)] {pat05} Patsourakos, S. \& Klimchuk, J. A. 2005, \apj, 628, 1023
	\bibitem[Peter et al.(2006)]            {pet06} Peter, H., Gudiksen, B. V., \& Nordlund, A. 2006, \apj, 638, 1086
	\bibitem[Porter \& Klimchuk(1995)]      {por95} Porter, L. J. \& Klimchuk, J. A. 1995, \apj, 454, 499
	\bibitem[Reale et al.(2005)]            {rea05} Reale, F., Nigro, G., Malara, F., Peres, G., \& Veltri, P. 2005, \apj, 633, 489
	\bibitem[Rosner et al.(1978)]           {ros78} Rosner, R., Tucker, W. H., \& Vaiana, G. S. 1978, \apj, 220, 643
	\bibitem[Schmelz et al.(2009)]          {sch09} Schmelz, J. T., Nasraoui, K., Rightmire, L. A., Kimble, J. A., Del Zanna, G., Cirtain, J. W., DeLuca, E. E., \& Mason, H. E. 2009, \apj, 691, 503
	\bibitem[Serio et al.(1981)]            {ser81} Serio, S., Peres, G., Vaiana, G. S., Golub, L., \& Rosner, R. 1981, \apj, 243, 288
	\bibitem[Serio et al.(1991)]            {ser91} Serio, S., Reale, F., Jakimiec, J., Sylwester, B., \& Sylwester, J. 1991, \aap, 241, 197
	\bibitem[Spadaro et al.(2003)]          {spa03} Spadaro, D., Lanza, A. F., Lanzafame, A. C., Karpen, J. T., Antiochos, S. K., Klimchuk, J. A., \& MacNeice, P. J. 2003, \apj, 582, 486
	\bibitem[Testa et al.(2005)]		        {tes05} Testa, P., Peres, G., \& Reale, F. 2005, \apj, 622, 695
	\bibitem[Walsh et al.(1997)]		        {wal97} Walsh, R.W., Bell, G.E., \& Hood, A.W. 1997, \solphys, 171, 81
	\bibitem[Winebarger et al.(2003)]       {win03} Winebarger, A. R., Warren, H. P., \& Mariska, J. T. 2003, \apj, 587, 439
\end{thebibliography}
\end{document}